# Improving End-To-End Modeling for Mispronunciation Detection with Effective Augmentation Mechanisms

Tien-Hong Lo, Yao-Ting Sung and Berlin Chen
National Taiwan Normal University, Taipei City, Taiwan
{teinhonglo, sungtc, berlin}@ntnu.edu.tw

*Abstract*— Recently, end-to-end (E2E) models, which allow to take spectral vector sequences of L2 (second-language) learners' utterances as input and produce the corresponding phone-level sequences as output, have attracted much research attention in developing mispronunciation detection (MD) systems. However, due to the lack of sufficient labeled speech data of L2 speakers for model estimation, E2E MD models are prone to overfitting in relation to conventional ones that are built on DNN-HMM acoustic models. To alleviate this critical issue, we in this paper propose two modeling strategies to enhance the discrimination capability of E2E MD models, each of which can implicitly leverage the phonetic and phonological traits encoded in a pretrained acoustic model and contained within reference transcripts of the training data, respectively. The first one is input augmentation, which aims to distill knowledge about phonetic discrimination from a DNN-HMM acoustic model. The second one is label augmentation, which manages to capture more phonological patterns from the transcripts of training data. A series of empirical experiments conducted on the L2-ARCTIC English dataset seem to confirm the efficacy of our E2E MD model when compared to some top-of-the-line E2E MD models and a classic pronunciation-scoring based method built on a DNN-HMM acoustic model.

## I. INTRODUCTION

With the ever-increasing surge of learning foreign languages and the continued progress of automatic speech recognition (ASR), computer-assisted pronunciation training (CAPT) has established itself as an integral means to alleviate the lack of qualified teachers and offer an individualized and self-paced environment for second-language (L2) learners to improve their speaking proficiency [1, 2]. Essential to the success of a CAPT system is the accuracy of the mispronunciation detection (MD) module, which is concerned with the identification of erroneous pronunciations in the utterance of an L2 learner that differ from the canonical pronunciations of the corresponding text prompt presented to the learner.

There have been substantial research efforts put into the development of effective methods to detect phone-level mispronunciation patterns of L2 learners by leveraging ASR-related techniques, which can be broadly grouped into two categories. The first category of methods is pronunciation-scoring based ones [3, 4, 5, 6], among which Goodness of Pronunciation (GOP) and its variants are the best-known instantiations [7]. GOP computes the ratio between the likelihoods of the canonical and the most likely pronounced phones predicted by an acoustic model, which is obtained through forced-alignment of the canonical phone sequence of a given text prompt with the speech signal uttered by a learner. A phone segment is identified as a mispronunciation if the corresponding ratio falls below a given threshold. GOP has been empirically shown to correlates well with human assessments. The other category of methods makes the end-to-end (E2E) ASR paradigm straightforwardly applicable to mispronunciation detection. A common practice of these E2E methods is to first employ a free-phone recognition process [8-14], implemented with deep neural networks like connectionist temporal classification (CTC) [15], attention-based model [16-18] or their hybrids (denoted by CTC-ATT for short) [19], to dictate the possible sequence of phones produced by an L2 learner. Then, the recognition transcript in turn can be compared to the canonical phone sequence of a text prompt to simultaneously detect and give feedback on mispronunciations. These E2E methods have shown promising mispronunciation detection performance, in relation to variants of the GOP-based method built on hybrid DNN-HMM acoustic models.

However, the models of current E2E neural methods for mispronounce detection are more likely to overfit due to huge numbers of parameters often need to be estimated and the lack of sufficient labeled speech data of L2 speakers, since collecting and transcribing an adequate amount of L2 speech data is often expensive and time-consuming. In view of this crucial issue, in this paper we make attempts to develop two disparate modeling strategies to enhance the discrimination capability of E2E MD models, each of which can implicitly leverage the phonetic and phonological traits encoded in a pretrained acoustic model and contained within reference transcripts of the training data, respectively. The first one is input augmentation, which aims to distill knowledge about phonetic discrimination (in the form of phonetic posteriorgram, PPG) from a hybrid DNN-HMM acoustic model. The second one is label augmentation, which manages to capture more phonological patterns from the transcripts of training data pooled from both L1 (first-language) and L2 utterances [20, 21, 22]. Extensive experiments carried out on a benchmark English dataset appear to demonstrate the utility of our modeling strategies.

The rest of this paper is organized as follows. We introduce the general procedure of the E2E-based approach to MD in Section 2. Section 3 elucidates how we perform input augmentation and label augmentation for use in the training of E2E MD models. After that, the experimental settings and extensive sets of MD experiments are presented in Sections 4 and 5, respectively. Finally, Section 6 concludes the paper with a remark on the future prospects of research.

## II. MODEL ARCHITECHTURE

### A. CTC Modeling Component

The formulation of CTC directly follows from the Bayes decision rule, viz. finding an output symbol (letter, phone or word-piece) sequence $C = c_1, c_2, \ldots c_L$ that has the maximum posterior probability given the frame-wise acoustic vector sequence $\mathbf{X} = \mathbf{x}_1 \mathbf{x}_2, \ldots, \mathbf{x}_T$ of an input utterance, which can be further factorized as follows:

$$P_{\text{CTC}}(C|\mathbf{X}) \approx \sum_S P(C|S) P(S|\mathbf{X}), \quad (1)$$

where $S = s_1, s_2, \ldots, s_T$ is a frame-wise, latent symbol sequence, while $P(S|\mathbf{X})$ and $P(C|S)$ are referred to as the acoustic model (or encoder) and translation model (or decoder) of CTC, respectively [15]. The acoustic model $P(S|\mathbf{X})$ can be instantiated with deep neural networks such as bi-directional long short-term memory (BLSTM), Transformer, and among others. Note also here that the logit $\mathbf{h}_t$ corresponding to each frame-wise latent symbol $s_t$ can be regarded as a distilled embedding vector that encodes the acoustic characteristics of the input $\mathbf{X}$ at time $t$. On a separate front, the translation model can be embodied with letter (alternatively, phone or word-piece) based language model with the first-order Markov assumption.

### B. Attention-based Modeling Component

The attention-based model outputs an output symbol sequence $C$ that has the maximum probability given an input acoustic vector sequence $\mathbf{X}$, without making any independence assumption on the output symbol sequence:

$$P_{\text{ATT}}(C|\mathbf{X}) = \prod_{l=1}^{L} P(c_l|\mathbf{X}, c_{1:l-1}), \quad (2)$$

where the posterior probability $P(c_l|\mathbf{X}, c_{1:l-1})$ can be estimated with an encoder working jointly with a decoder. One thing to note is that both CTC and the attention-based model can employ the same model formulation for their encoders.

### C. Hybrid CTC-Attention Approach

Since the attention-based model has the disadvantages of non-monotonous left-to-right alignment and slow convergence, whereas CTC has to use an additional well-trained translation (language) model to achieve better results, there is good reason to integrate them together [19], especially leveraging CTC to help constrain the left-to-right alignment order between the input and output sequences when decoding a possible output sequence:

$$P_{\text{CTC-ATT}}(C|\mathbf{X}) = \lambda P_{\text{CTC}}(C|\mathbf{X}) + (1-\lambda) P_{\text{ATT}}(C|\mathbf{X}) \quad (3)$$

where $\lambda$ a non-negative parameter used to control the relative importance of $P_{\text{CTC}}(C|\mathbf{X})$ and $P_{\text{ATT}}(C|\mathbf{X})$. As aforementioned in Section I, when the CTC-ATT model is applied to CAPT, the recognition transcript output from CTC-ATT in response to a learner's utterance can be directly compared to the canonical phone sequence of the corresponding text prompt to simultaneously detect and give feedback on mispronunciations.

## III. PROPOSED APPROACH

Despite E2E neural methods have recently shown some success as a promising alternative to the classic GOP-based method and its variants for mispronunciation detection, their models are still prone to overfitting due to a huge number of parameters often need to be estimated for an E2E neural method and the possible lack of sufficient labeled training data. In this section, we describe in detail two simple yet effective mechanisms proposed in this paper to work around the aforementioned issues.

The first is to perform input augmentation (IA) for a CTC-ATT MD model by concatenating the original input spectral feature vector of each speech frame with a phonetic posteriorgram (PPG) vector which is produced by a pretrained DNN-HMM acoustic model. PPG generally can be referred to as the posterior probability that each speech frame belongs to a set of predefined phonetic units. Further, we also can view the corresponding PPG vectors of an input speech utterance collectively as the knowledge of the acoustic model capture from the utterance [23], which is expected to retain the linguistic and phonetic discrimination information of the utterance [24, 25]. In doing such kind of input augmentation, we can not only distill knowledge about linguistic and phonetic discrimination from the frame-level output of a hybrid DNN-HMM acoustic model, but to some extent make the input to the E2E MD model less affected by a wide variety of subtle factors such as speaker, gender, age, accent, channel, and among others, which a CAPT system is often confronted with. In this work, the hybrid DNN-HMM acoustic model was trained on both the L1 (viz. the TIMIT dataset [26]) and the L2 (viz. the L2-ARCTIC dataset [27]) speech corpora and in turn used to extract the phonetic PPG vector that corresponds to each speech fame of an L2 learning's utterance. The notion of leveraging PPG-related information to replace or augment spectral features has been recently investigated voice conversion [24] and cross-accent voice conversion [25]. However, to our best knowledge, this notion remains relatively underexplored in the context of E2E-based mispronunciation detection.

On a separate front, apart from widely-adopted techniques like dropout, early stopping and batch normalization, label smoothing (or output regularization) has emerged as a promising therapy to prevent E2E neural models from overfitting especially when there is a discrepancy between training and test conditions [20, 21, 22]. The model-agnostic property of label smoothing (viz. invariant to the parameterization of E2E neural models) makes it broadly applicable to image classification and speech recognition, just to name a few. We thus follow common practice to train our baseline E2E MD model with label smoothing which is instantiated with the so-called unigram smoothing mechanism [21]. Though other variants of label smoothing, such as uniform smoothing [20] and temporal smoothing [22], are also applicable for the purpose, unigram smoothing is employed here mainly due to that it can be exploited to capture the phonological patterns resided in the transcripts of training data. To flesh this out a bit more, the training objective of the baseline E2E MD model, normally in the form of negative log-likelihood, is incorporated with a penalty term that measures the KL divergence between a unigram distribution $\mathcal{D}_{\text{Uni}}$ and the predicted distribution of the E2E MD model $\mathcal{D}_{\text{CTC-ATT}}(\mathbf{X})$ for each training utterance $\mathbf{X}$:

$$\mathcal{L}(\theta) = \sum_{\mathbf{X} \in \text{Traing Set}} -\log P_{\text{CTC-ATT}}(C|\mathbf{X}) + \alpha \cdot D_{KL}(\mathcal{D}_{\text{Uni}} || \mathcal{D}_{\text{CTC-ATT}}(\mathbf{X})), \quad (4)$$

where $\alpha$ control the strength of the penalty term and $\mathcal{D}_{\text{Uni}}$ is estimated based on the transcripts of L1 and L2 training data, and $C$ is the corresponding orthographic transcript of $\mathbf{X}$.

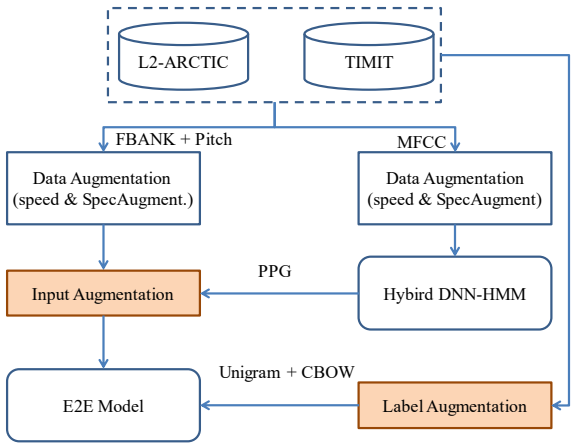

Fig. 1 A schematic depiction of our proposed modeling approach to mispronunciation detection.

Since the estimation of $\mathcal{D}_{\text{Uni}}$ merely depends on the raw counts of phonetic units occurring in the training data, it may not render well the phonological patterns manifested in the transcripts of training data. To make unigram smoothing more effective, we leverage the Continuous Bag of Words (CBOW) method [28] to seize the intrinsic temporal structure (with respect to phonological patterns) inherent in the transcripts of training data. For the idea to work, a unigram probability distribution estimated with CBOW is accomplished through the following equations:

$$\mathbf{v}_{phn_i} = CBOW(phn_i), \quad (5)$$

$$\mathbf{u}_{phn_i} = [cos(\mathbf{v}_{phn_1}, \mathbf{v}_{phn_i}), \dots, cos(\mathbf{v}_{phn_I}, \mathbf{v}_{phn_i})], \quad (6)$$

$$\mathcal{D}_{\text{CBOW}}(phn_i) = softmax(\|\mathbf{u}_{phn_i}\|), \quad (7)$$

where $\mathbf{v}_{phn_i}$ is the vector representation of a phone unit $i$ generated by CBOW; $cos(\mathbf{v}_{phn_i}, \mathbf{v}_{phn_j})$ is the cosine similarity between phone units $i$ and $j$; and $I$ is the total number of distinct phone units. As such, we can linearly interpolate $\mathcal{D}_{\text{Uni}}$ with $\mathcal{D}_{\text{CBOW}}$ to form a new unigram distribution for enhanced label smoothing:

$$\widetilde{\mathcal{D}}_{\text{Uni}} = \beta \cdot \mathcal{D}_{\text{CBOW}} + (1 - \beta) \cdot \mathcal{D}_{\text{Uni}}, \quad (8)$$

where $\beta$ is introduced to specify the relative importance of the contextual distribution $\mathcal{D}_{\text{CBOW}}$ and the original unigram $\mathcal{D}_{\text{Uni}}$. Henceforth we will refer to such an enhanced label smoothing approach as the label augmentation (LA) mechanism. Figure 1 shows a schematic illustration of our proposed modeling approach to mispronunciation detection.

Table 1. Details of corpus used in our experiments.

|  |  | Spks. | Utts. | Hrs. |
|---|---|---|---|---|
| TIMIT | Train | 462 | 3,696 | 3.15 |
|  | Dev | 50 | 400 | 0.34 |
|  | Test | 24 | 192 | 0.16 |
| L2-ARCTIC | Train | 17 | 2,549 | 2.66 |
|  | Dev | 1 | 150 | 0.12 |
|  | Test | 6 | 900 | 0.88 |

IV. EXPERIMENTAL SETUP

We present in this section the experimental datasets and the metrics we employed to evaluate our proposed approach to mispronunciation detection.

A. Speech Corpora and Evaluation Metrics

A series of mispronunciation detection experiments were conducted the L2-ARCTIC benchmark corpus [27]. L2-ARCTIC is an open-access L2-English speech corpus compiled for research on CAPT, accent conversion, and others. About 3,600 utterances of 24 non-native speakers (12 males and 12 females; equipped with manual transcripts) across different nationalities, some of which contained mispronounces, collectively constitute L2-ARCTIC (Version 4). These speakers are made up of six mother-tongue languages, including Hindi, Korean, Mandarin, Spanish, Arabic and Vietnamese. In addition, a limited amount of native (L1) English speech data compiled from the TIMIT corpus [26] (composed of 630 speakers) was pooled together with the training set of L2-ARCTIC for the estimation of the various

Table 2. Results of our various E2E models on correct pronunciation and mispronunciation detection, in comparison to some state-of-the art models.

| | Input Augmentation (IA) | Label Augmentation (LA) | L2-ARCTIC | | | | | | TIMIT |
|---|---|---|---|---|---|---|---|---|---|
| | | | Correct Pronunciation Detection (CD) | | | Mispronunciation Detection (MD) | | | PER (%) |
| | | | RE (%) | PR (%) | F1 (%) | RE (%) | PR (%) | F1 (%) | |
| GOP | - | - | 90.98 | 91.97 | 91.47 | 50.15 | 46.99 | 48.52 | - |
| CNN-RNN-CTC | - | - | 79.97 | 93.88 | 86.37 | 67.29 | 34.88 | 45.94 | 18.15 |
| CTC-ATT | - | - | 87.47 | 93.40 | 90.34 | 61.23 | 43.80 | 51.07 | 15.50 |
| | V | - | 91.40 | 92.85 | 92.12 | 55.88 | 50.89 | 53.26 | 13.68 |
| | - | V | 87.82 | 93.33 | 90.49 | 60.64 | 44.25 | 51.16 | 14.73 |
| | V | V | 90.77 | 92.82 | 91.78 | 55.98 | 49.15 | 52.35 | 14.03 |
| CTC-ATT (SpecAugment) | - | - | 91.44 | 91.61 | 91.52 | 47.76 | 46.92 | 47.19 | 16.60 |
| | V | - | 92.63 | 92.44 | 92.53 | 52.47 | 53.17 | 52.81 | 13.14 |
| | - | V | 92.84 | 91.93 | 92.38 | 48.88 | 52.11 | 50.44 | 14.71 |
| | V | V | 92.83 | 92.37 | 92.60 | 51.90 | 53.57 | 52.72 | 13.31 |

E2E MDD models and the DNN-HMM acoustic model of the GOP-based method. To follow the setup of TIMIT [29], at training time, these models all employed an inventory of 48 distinct canonical phones, each of which was supplemented with a peculiar "anti-phone" for the purpose of labeling its corresponding mispronounced segments in L2-ARCTIC that were caused by non-categorical or distortion errors, viz. approximating L2 phones with L1 (first-language) phones, or erroneous pronunciation patterns in between. At test time, the original canonical phone inventory was consolidated into a new inventory of 39 distinct canonical phones, so as to make our experimental setup in line with previous work [11, 12, 29]. Furthermore, the setting of the mispronunciation detection experiments on L2-ARCTIC followed the recipe provided by [12]. The details of TIMIT and L2-ARCTIC corpus are shown in Table 1.

The default evaluation metric employed in this paper for the mispronunciation detection task is the F1-score, which is a harmonic mean of precision and recall, defined as:

$$F1 = \frac{2 * Precision * Recall}{Precision + Recall}, \quad (9)$$

$$Precison = \frac{C_{D \cap H}}{C_D}, \quad (10)$$

$$Recall = \frac{C_{D \cap H}}{C_H}, \quad (11)$$

where $C_D$ is the total number of phone segments in the test set that were identified as being mispronounced (or correct) by the current mispronunciation detection module, $C_H$ is the total number of phone segments in the test set that were identified as being mispronounced (or correct) by the human assessor, and $C_{D \cap H}$ were is the total number of phone segments in the test set that are identified as being mispronounced simultaneously by both the current mispronunciation detection module and the annotation of human assessors.

*B. Implementation Details*

Our baseline mispronunciation detection method was implemented with the CTC-Attention neural model and the ESPnet toolkit [30], while the GOP-based method with the DNN-HMM acoustic model was built on the Kaldi toolkit [31]. The default acoustic features fed into the CTC-Attention model were composed of 80-dimensional Mel-filter-bank feature vectors (FBANK) appended with 3-dimensional pitch features, while 40-dimensional Mel Frequency Cepstral Coefficients (MFCC) augmented with an i-vector of 100-dimensions for the DNN-HMM acoustic model [31]. In addition, the CTC-Attention neural model of our baseline E2E method is implemented with VGG-BLSTM for the shared encoder and two-layer LSTM for the decoder of the attention-based model. Meanwhile, the DNN-HMM acoustic model was instantiated with CNN-TDNNF [32]. When training our baseline E2E model and the DNN-HMM acoustic model of the GOP-based method, the commonly-used data augmentation method (viz. SpecAugment [33]) was optionally applied to augment the training set, with the hope for better model estimation.

Table 3. The impacts of using different input feature vectors (viz. FBANK and PPG) and their concatenation on mispronunciation detection

| Model | Input Feature Vectors | Mispronunciation Detection (MD) | | |
|---|---|---|---|---|
| | | RE (%) | PR (%) | F1 (%) |
| CTC-ATT | FBANK | 61.23 | 43.80 | 51.07 |
| | PPG | 55.18 | 48.97 | 51.89 |
| | FBANK+PPG | 55.88 | 50.89 | 53.26 |

On the other hand, for the label augmentation mechanism described in Section III, the CBOW-derived vector representations of canonical phones were estimated based on the training set, while the hyperparameters $\alpha$ (*cf.* Eq. (4)) and $\beta$ (*cf.* Eq. (8)) were both set to be 0.1 in this study.

## V. EXPERIMENTAL RESULTS

### A. Baseline Results

In the first place, we assess the performance level of our baseline E2E model (denoted by CTC-ATT; *cf.* Section II) trained with the default setting of label smoothing (*cf.* Eq. (4)), in relation to the celebrated GOP-based model [7] and the recently proposed E2E model with the so-called CNN-RNN-CTC neural structure [10]. The corresponding results are shown in Table 2. We can observe from Table 2 that our baseline model (*cf.* Row 3) outperforms GOP (*cf.* Row 1) on mispronunciation detection (MD), whereas the situation is reversed for correct pronunciation detection (CD). CNN-RNN-CTC, however, performs the worst among them.

### B. Evaluations on Proposed Augmentation Mechanisms

We now turn to evaluate the efficacy of our two augmentation mechanisms, viz. input augmentation (IA) and label augmentation (LA) in enhancing the MD and CD performance of our E2E model. As can be seen from Rows 4 to 6 of Table 2, input augmentation (*cf.* Row 4) can improve the MD and CD performance of our model greatly, while the improvements brought by label augmentation (*cf.* Row 5) is not as significant as input augmentation. One possible reason is that since our baseline system had already trained with label smoothing (*cf.* Eq. (4)), label augmentation (*cf.* Eq. (8)) can be regarded as an enhanced treatment of label smoothing and thus offers less pronounced gains. On the other hand, the synergy of these two augmentation mechanisms (*cf.* Row 6) does not provide additional performance gains with respect to MD and CD; further investigation would still be desirable. In addition, we also evaluate effectiveness of the renowned augmentation mechanism SpecAugment, which were widely-adopted for E2E ASR, in the context of CAPT. It is evident from the last four rows of Table 2, when SpecAugment is applied to model training in isolation (*cf.* Row 7), it results in a notable improvement over our baseline E2E model for CD but deteriorated performance for MD. Further, if we combine SpecAugment with either one/both of our augmentation mechanisms, only a mild improvement CD is observed. To recap here, our input augmentation mechanism, which augments the original spectral feature vectors (viz. FBANK) with the phonetic posteriorgram (PPG) vectors produced by the pretrained DNN-HMM acoustic model, seems to hold promise for the CAPT task, yielding marked performance improvements over GOP and CNN-RNN-CTC for both MD and CD. Further, the phone error rate (PER) result of ASR on the test set of TIMIT (*cf.* the rightmost column of Table 2) can also be considerable promoted when our best-performing model is alternatively employed for this purpose.

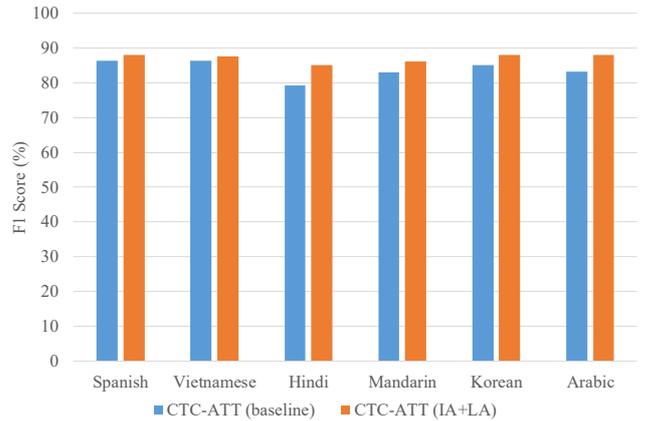

Fig. 2 The mispronunciation detection performance of our proposed augmentation mechanisms for test speakers of different mother-tongue languages.

Table 4. The ten confusing phone pairs that most frequently occur in the test set of L2-ARCTIC due to mispronunciations.

| | | | Model | |
|---|---|---|---|---|
| | | | CTC-ATT (IA+LA) | |
| Err. | Type | Count | Detection Accuracy Rate (%) | Diagnosis Accuracy Rate (%) |
| dh->d | Sub. | 436 | 64.68 | 89.01 |
| z->s | Sub. | 397 | 57.93 | 96.52 |
| ih->iy | Sub. | 159 | 28.93 | 95.65 |
| d->SIL | Del. | 145 | 50.34 | 79.45 |
| t->SIL | Del. | 136 | 47.06 | 87.50 |
| r->SIL | Del. | 135 | 75.56 | 83.33 |
| ow->aa | Sub. | 116 | 15.52 | 88.89 |
| er->ah | Sub. | 98 | 65.31 | 75.00 |
| d->t | Sub. | 73 | 23.29 | 70.59 |
| l->SIL | Del. | 63 | 77.78 | 59.18 |

### C. Ablation Studies

We then move on to conduct two ablation studies to analyze the utility of our models from different aspects. The first one is to understand the impacts of using different input feature vectors (viz. FBANK and PPG) and their concatenation on mispronunciation detection (MD). The corresponding results are shown in Table 3, where the results of FBANK (the baseline) and FBANK+PPG (viz. merely with input augmentation) are directly adopted from Row 3 and Row 4 of Table 2, respectively. We see that FBANK delivers a higher recall rate (RE) than PPG for the MD purpose, whereas the later provides better precision rate (PR) and F1-score results than the former. The combination of them can bring salient improvements in terms of the precision rate and F1-score metrics, while the improvement for the recall rate is marginally significant. The second ablation study is to investigate to what extent the combination of our two augmentation mechanisms can improve the MD performance (F1-score) for the test speakers respectively of each mother-tongue language. As it is evident from Figure 2, the MD performance improvements for the Hindi and Arabic speakers are more obvious than the speakers of other mother-tongue languages. In particular, as pointed out in [27], Hindi speakers often use segmental and suprasegmental features in ways that are distinct from American speakers or non-native speakers of other mother-tongue languages.

As a final note, we summarize in Table 4 the ten confusing phone pairs that most frequently occur in the test set of L2-ARCTIC due to mispronunciations (e.g., dh->d means phone unit dh is mispronounced as phone unit d), where their corresponding detection and diagnosis accuracy rates [4, 11, 12] are also listed for reference. Apart from the phone deletion errors, it seems that most of the substitution errors could be properly detected if some duration cues of phonetic pronunciations are further fused into our proposed neural MD models.

## VI. Conclusion and Future Work

In this paper, we have presented a novel E2E modeling paradigm for mispronunciation detection (MD). In particular, two simple yet feasible augmentation mechanisms, viz. input augmentation and label augmentation, have also been proposed. Extensive sets of experiments conducted on the L2-ARCTIC benchmark dataset seem to show the effectiveness and practical feasibility of our modeling paradigm, in comparison to some top-of-the-line models. In the future, we would like to explore more fine-grained acoustic, prosodic, accent or suprasegmental pronunciation phenomena [35], as well as practically unfeasible ways to integrate them into our modeling paradigm.

## VII. Acknowledgement

This research is supported in part by Ministry of Science and Technology, Taiwan under Grant Number MOST 110-2634-F-008-004- through Pervasive Artificial Intelligence Research (PAIR) Labs, Taiwan, and Grant Numbers MOST 108-2221-E-003-005-MY3 and MOST 109-2221-E-003-020-MY3. Any findings and implications in the paper do not necessarily reflect those of the sponsors.

## References

[1] A. Neri, O. Mich, M. Gerosa and D. Giuliani, "The effectiveness of computer assisted pronunciation training for foreign language learning by children," in *Computer Assisted Language Learning*, vol. 21, no. 5, pp. 393–408, 2008.

[2] C. Tejedor-García, D. Escudero-Mancebo, E. Cámara-Arenas, C. González-Ferreras and V. Cardeñoso-Payo, "Assessing pronunciation improvement in students of english using a controlled computer-assisted pronunciation tool," in *IEEE Transactions on Learning Technologies*, vol. 13, no. 2, pp. 269-282, 2020

[3] H. Li, S. Huang, S. Wang and B. Xu, "Context-dependent duration modeling with backoff strategy and look-up tables for pronunciation assessment and mispronunciation detection," in Proc. *Interspeech*, 2011, pp. 1133–1136.

[4] K. Li, X. Qian and H. Meng, "Mispronunciation Detection and Diagnosis in L2 English Speech Using Multidistribution Deep Neural Networks," in *IEEE/ACM Transactions on Audio, Speech, and Language Processing*, vol. 25, no. 1, pp. 193-207, 2017.

[5] S. Sudhakara, M. K. Ramanathi, C. Yarra and P. K. Ghosh, "An improved goodness of pronunciation (GOP) measure for pronunciation evaluation with dnn-hmm system considering hmm transition probabilities." in *Proc. Interspeech*, 2019, pp. 954–958.

[6] S. Cheng, Z. Liu, L. Li, Z. Tang, D. Wang and T. F. Zheng, "Asr-free pronunciation assessment," in *arXiv preprint arXiv:2005.11902*, 2020.

[7] S. M. Witt and S. J. Young, "Phone-level pronunciation scoring and assessment for interactive language learning," *Speech communication*, vol. 30, no. 2-3, pp. 95–108, 2000.

[8] N. Minematsu, "Pronunciation assessment based upon the phonological distortions observed in language learners' utterances," in *Proc. Interspeech*, 2004.

[9] A. Lee and J. R. Glass, "Pronunciation assessment via a comparison-based system," in *Proc. SLaTE*, 2013.

[10] W. Leung, X. Liu and H. Meng., "CNN-RNN-CTC Based End-to-end Mispronunciation Detection and Diagnosis," in *Proc. ICASSP*, pp. 8132-8136, 2019.

[11] K. Fu, J. Lin, D. Ke, Y. Xie, J. Zhang and B. Lin, "A Full Text-Dependent End to End Mispronunciation Detection and Diagnosis with Easy Data Augmentation Techniques." in *arXiv preprint arXiv:2104.08428*, 2021.

[12] Y. Feng, G. Fu, Q. Chen and K. Chen, "SED-MDD: Towards Sentence Dependent End-To-End Mispronunciation Detection and Diagnosis," in *Proc. ICASSP*, pp. 3492-3496, 2020.

[13] T.-H. Lo, S.-Y. Weng, H.-J. Chang and B. Chen, "An effective end-to-end modeling approach for mispronunciation detection," in *Proc. Interspeech*, pp. 3032–3036, 2020.

[14] B.-C. Yan, M. C. Wu, H. T. Hung and B. Chen, "An end-to-end mispronunciation detection system for L2 English speech leveraging novel anti-phone modeling," in *Proc. Interspeech*, pp. 3032–3036, 2020.

[15] A. Graves, S. Fernández, F. Gomez and J. Schmidhuber, "Connectionist temporal classification: labelling unsegmented sequence data with recurrent neural networks," in *Proc. ICML*, pp. 369-376, 2006.

[16] J. Chorowski, D. Bahdanau, K. Cho and Y. Bengio, "End-to-end continuous speech recognition using attention-based recurrent nn: First results," in *arXiv preprint arXiv:1412.1602*, 2014.

[17] J. Chorowski, D. Bahdanau, D. Serdyuk, K. Cho and Y. Bengio, "Attention-based models for speech recognition," in *Proc. NIPS*, pp. 577–585, 2015.

[18] D. Bahdanau, J. Chorowski, D. Serdyuk, P. Brakel and Y. Bengio, "End-to-end attention-based large vocabulary speech recognition," in *Proc. ICASSP*, pp. 4945–4949, 2016.

[19] S. Kim, T. Hori and S. Watanabe, "Joint CTC-Attention based end-to-end speech recognition using multi-task learning," in *Proc. ICASSP*, pp. 4835-4839, 2017.

[20] C. Szegedy, V. Vanhoucke, S. Ioffe, J. Shlens and Z. Wojna, "Rethinking the inception architecture for computer vision," in *Proc. CVPR*, pp. 2818-2826, 2016

[21] G. Pereyra, G. Tucker, J. Chorowski, Ł. Kaiser and G. Hinton, "Regularizing neural networks by penalizing confident output distributions," in *Proc. ICLR*, 2017.

[22] J. Chorowski and N. Jaitly, "Towards better decoding and language model integration in sequence to sequence models," in *Proc. Interspeech*, 2017.

[23] G. Hinton, O. Vinyals and J. Dean, "Distilling the knowledge in a neural network," in *arXiv preprint arXiv:1503.02531*, 2015.

[24] L. Sun, H. Wang, S. Kang, K. Li and H. M. Meng, "Personalized, cross-lingual tts using phonetic posteriorgrams," in *Proc. Interspeech*, pp. 322-326, 2016.

[25] G. Zhao, S. Ding and R. Gutierrez-Osuna, "Foreign accent conversion by synthesizing speech from phonetic posteriorgrams," in *Proc. Interspeech*, pp. 2843–2847, 2019.

[26] J. S. Garofolo, L. F. Lamel, W. M. Fisher, J. G. Fiscus and D. S. Pallett, "Darpa timit acoustic-phonetic continous speech corpus cd-rom. nist speech disc 1-1.1," in *NASA STI/Recon technical report n*, vol. 93, 1993.

[27] G. Zhao, S. Sonsaat, A. O. Silpachai, I. Lucic, E. Chukharev-Hudilainen, J. Levis and R. Gutierrez-Osuna, "L2-arctic: A non-native english speech corpus," *Perception Sensing Instrumentation Lab*, 2018.

[28] T. Mikolov, K. Chen, G. Corrado and J. Dean, "Efficient estimation of word representations in vector space.", in *arXiv preprint arXiv:1301.3781*, 2013.

[29] K-F Lee and H-W Hon, "Speaker-independent phone recognition using hidden markov models," in *IEEE Transactions on Acoustics, Speech, and Signal Processing*, vol. 37, no. 11, pp. 1641–1648, 1989.

[30] S. Watanabe, T. Hori, S. Karita, T. Hayashi, J. Nishitoba, Y. Unno, N. E. Y. Soplin, J. Heymann, M. Wiesner, N. Chen, A. Renduchintala and T. Ochiai, "ESPnet: end-to-end speech processing toolkit," in *Proc. Interspeech*, pp. 2207-2211, 2018.

[31] D. Povey, A. Ghoshal, G. Boulianne, L. Burget, O. Glembek, N. Goel, M. Hannemann, P. Motlíček, Y. Qian, P. Schwarz, J. Silovsky, G. Stemmer and K. Vesely, "The kaldi speech recognition Toolkit," in *Proc. ASRU*, 2011.

[32] N. Dehak, P. J. Kenny, R. Dehak, P. Dumouchel and P. Ouellet, "Front-end factor analysis for speaker verification," in *IEEE Transactions on Audio, Speech, and Language Processing*, vol. 19, no. 4, pp. 788–798, 2011.

[33] D. Povey, G. Cheng, Y. Wang, K. Li, H. Xu, M. Yarmohamadi and S. Khudanpur., "Semi-orthogonal low-rank matrix factorization for deep neural networks," in *Proc. Interspeech*, pp. 3743-3747, 2018.

[34] D. S. Park, W. Chan, Y. Zhang, C. C. Chiu, B. Zoph, E. D. Cubuk and Q. V. Le, "SpecAugment: A simple data augmentation method for automatic speech recognition", in *Proc. Interspeech*, pp. 2613-2617, 2019.

[35] S.-W. Fan Jiang, B.-C. Yan, T.-H. Lo, F.-A. Chao and B. Chen, "Towards robust mispronunciation detection and diagnosis for L2 English learners with accent-modulating methods," in *Proc. ASRU*, 2021.